# Temperature Dependence of a Sub-wavelength Compact Graphene Plasmon-Slot Modulator

Zhizhen Ma[1], Rubab Amin[1], Sikandar Khan[1], Mohammad Tahersima[1], Volker J. Sorger[1*]

[1]*Department of Electrical and Computer Engineering, George Washington University, Washington, D.C. 20052, USA*
*Author e-mail address: sorger@gwu.edu*

**Abstract:** We investigate a plasmonic electro-optic modulator with an extinction ratio exceeding 1 dB/μm by engineering the optical mode to be in-plane with the graphene layer, and show how lowering the operating temperature enables steeper switching. We show how cooling Graphene enables steeping thus improving dynamic energy consumption. Further, we show that multi-layer Graphene integrated with a plasmonic slot waveguide allows for in-plane electric field components, and 3-dB device lengths as short as several hundred nanometers only. Compact modulators approaching electronic length-scales pave a way for ultra-dense photonic integrated circuits with smallest footprints to-date, and picosecond short RC delay times, beneficial for fast modulation formats.

## I. INTRODUCTION

With device down-scaling the capacitive delay of electronics increasingly limits communication overhead. Photonics on the other hand bears a notion of parallelism since a) photons only interact through media with each other, and b) the bosonic character allow a state to occupy the same quantum state [1]. As a result, optics is therefore prone to be a candidate-of-choice for data communication. While the prospects for photonics are known, optoelectronic devices need either be footprint-density competitive to electronics or outperform electronics, in order to become dominant at the chip market. Loosely put, the expected break-even performance of diffraction limited opto-electronic devices relative to electronics has to be about 100 times to make up for the larger footprint (electronics = 20 nm, photonics diffraction limit = $\lambda/2n \sim 200$ nm, where $\lambda$ = telecom wavelength). This barrier-to-entry could be lowered by scaling-down opto-electronics beyond the diffraction limit, which is actually synergistic to device performance such as discussed here.

There are a variety of 2D materials, but for electro-optic modulation graphene's Pauli Blocking has shown decent functionality as demonstrated by ref [2]. Here we also consider graphene, however other 2D materials are similarly synergistic to the design rules applied here, i.e. in-plane field components of the optical mode. Indeed effort has been made in integrating graphene [3, 4] with plasmonics with the purpose of modulation [5-10]. Yet, the anisotropy of 2D films introduce challenges with respect to polarization alignment using plasmonics since field lines are always perpendicular to a metal plane. As a result, plasmonic approaches, thus far, have shown low modulation capability and non-synergistic device designs despite graphene's strong index modulation potential [4]. Nevertheless, graphene phase modulation shows tens of GHz fast modulation, however relies on the strong feedback from a mirroring cavity leading to non-compact footprints and temperature sensitivities [11]. Thus, in this study we focus on engineering the optical mode profile of graphene to enhance the light-matter interaction while using a plasmonic modulation platform to decrease the device footprint. Previously we showed that volumetric device shrinking (scaling) of EO modulators fundamentally enables to reduce the energy consumption (E/bit), and sub fJ/bit are possible for plasmonic modulator design [12].

## II. METHOD

Graphene is an anisotropic material given its dimensions: in its honeycomb like lattice plane, the in-plane permittivity ($\varepsilon_\parallel$) can be tuned by varying its chemical potential $\mu_c$, whereas the out-of-plane permittivity is reported to remain constant around 2.5. In this work, graphene is modeled in two different temperatures by Kubo model at T = 0K and random phase approximation (RPA) at T = 300K (Fig. 1). We note, that the drastic change in graphene's imaginary refractive index, $\kappa$, is due to the strong effect through Pauli blocking, thus making graphene naturally suitable material for electro-absorption modulators. At higher temperature, the imaginary refractive index vs. chemical potential is smeared due to the natural temperature dependency of the Fermi-Dirac distribution function, leading to a sharp transition upon cooling. Doping (i.e. biasing) graphene to a chemical potential near half of the photon's energy, a small switching energy is needed for of only >0.05eV for T = 0K versus ~0.2eV at T = 300K (Fig. 1).

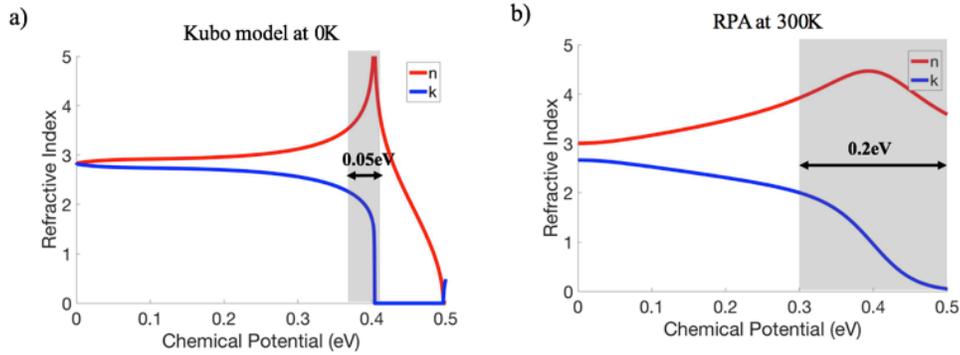

**Figure 1.** Refractive index plotted by Kubo (T = 0K) and RPA (T = 300K) model, where the black arrow shows the minimum chemical potential tuning needed for switching. a) Kubo model, b) RPA model. $\lambda_{photon}$ = 1550 nm (0.8eV).

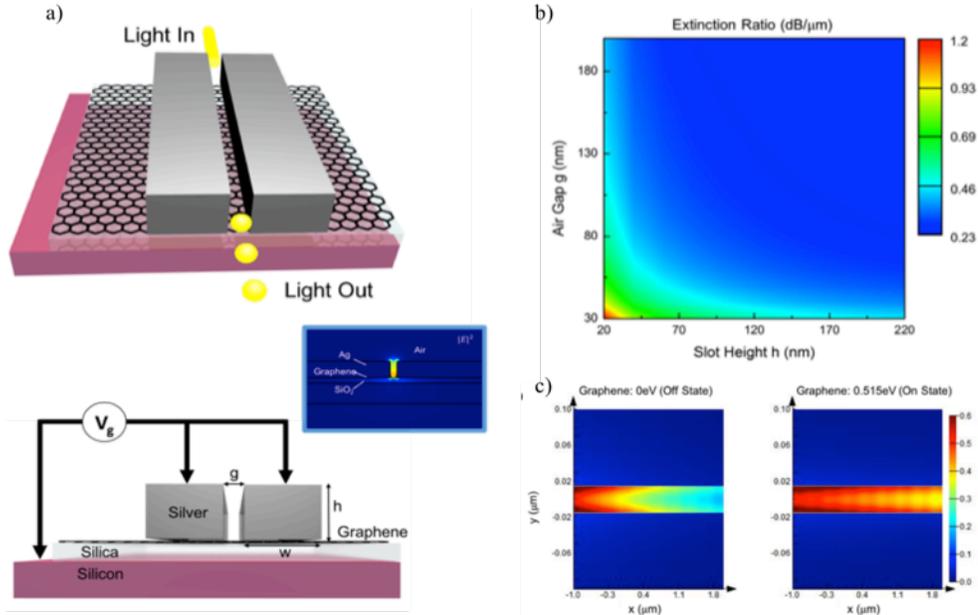

**Fig. 2. a)** The device schematic and the mode profile in the plasmonic slot as shown in the inset. **b)** The extinction ratio for different slot width and height, smaller slot dimension shows a better confinement where a higher ER is observed. **c)** Top view from the FDTD simulation, the power transmission at on/off state.

Our field-synergistic graphene modulator is sandwiched in between a plasmonic slot waveguide structure and silicon back gate (Fig. 2). Design optimization shows the expected higher extinction ratio upon improving the optical mode overlap more with the active graphene layer, here achieved by scaling the slot width and height to below 50 nm (Fig. 2b).

## III. RESULTS

For such a field-in-plane graphene modulator we find an extinction ratio (ER) of 1.2 dB/μm, which is ~10x higher compared to the highest previous-reported devices [13]. We previously showed an ER vs. overlap factor relation finding a 10x higher ER for a graphene modulator using plasmonic in-plane fields, indicating the synergistic integration potential of 2D materials with plasmonic waveguides [3].

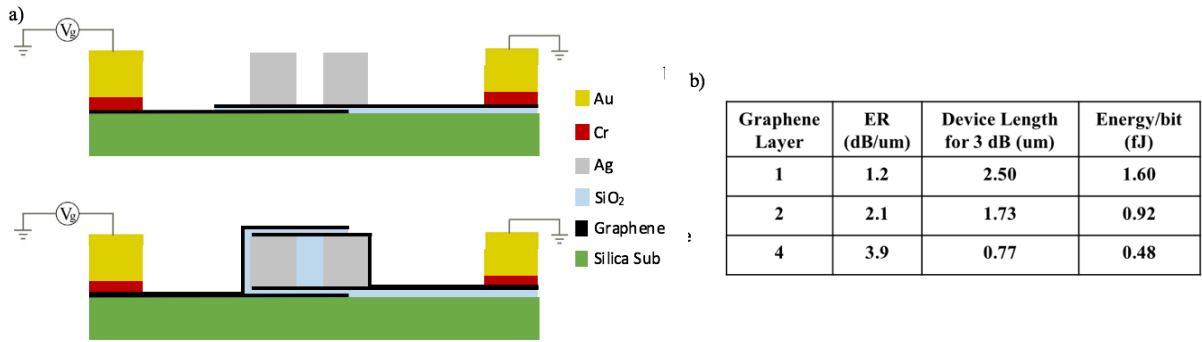

Figure 3. a) Graphene plasmonic slot EAM performance with different layers of 2 and 4. A thin oxide layer separates the graphene layer and every layer film interference wave guide structures. The ER increases almost linearly with increasing graphene layer, and energy consumption drops with decreasing device dimension.

Since graphene's Pauli blocking is symmetric to the applied bias (positive and negative potentials result equal responses) one can bias one graphene layer against another, similar to a push-pull configuration of MZIs [13]. Here we extend this configuration to 2 and 4 graphene layers, where in the latter one pair is place below and another on-top of the metallic slot. Our results show a near linear ER improvement with increasing the number or graphene layers (Fig. 3). This can understood by the added graphene layers a) being in close proximity to each other, and b) each pair having a similar interaction with the field of the slot waveguide. A small asymmetry originates from substrate effects of the modal profile (inset Fig. 2a). Remarkable for 2 graphene pairs nearly ER = 4 dB/μm is achieved resulting in a short device length of 0.77 μm for 3 dB modulation. Such compact device reduces the electrical capacitance, then enabling efficient device operation of just 500 aJ/bit at room temperature, whereas 100 aJ/bit are expected upon cooling the device to cryogenic temperatures for energy sensitive applications (Fig. 3b). However, due to the limited field leakage outside the plasmonic slot, the increase in ER is not significant beyond more than 4 layers of graphene.

In conclusion, graphene slot waveguide electro-absorption modulators with sub-λ long length scales are possible that preserve signal quality by providing high extinction ratios per nominal lengths (4 dB/μm), enabled by optimizing the in-plane field overlap of the active 2D material (graphene) in a plasmonic slot waveguide mode. The switching steepness improves upon thermal cooling allowing a reduction of the drive voltage to result in efficiencies approaching 100 aJ/bit. Such ultra-compact modulators enable a path for a class of opto-electronic devices approaching length scales just 10x of electronic device lengths. This allows for smallest capacitance beneficial from both energy and AC modulation point of view. Anticipated down scaling of photonic components for interconnect communication links [15,16] can lead to high-functional and reconfigurable network-on-chip solutions [16], to solve the I/O bottleneck in computing [17] and enabling cognitive compute architectures [18,19].

Architecture Letters, vol. 15, no. 1, pp. 1-4, (2016).

[19] V.S. is supported by ARO (W911NF-16-2- 0194) and by AFOSR (FA9550-14-1-0215).